\magnification\magstep 1
\baselineskip=0.70 true cm
\parindent = 0 true cm

\def\hor{\hskip 1.0 true cm}

\vglue 0.5 true cm
\vskip 1.0 true cm

\centerline{\bf NONLINEARITIES AND POMERON 
NONFACTORIZABILITY }
\centerline{\bf IN CONVENTIONAL DIFFRACTION}
\bigskip

\centerline{\bf P. Desgrolard$^{(1)}$,
A.I. Lengyel$^{(2)}$, E.S. Martynov$^{(3)}$.}

\bigskip

\centerline{$^{(1)}$ Institut de Physique Nucl\'eaire de 
Lyon, IN2P3-CNRS
et Universit\'e  Claude Bernard }
\centerline{ F 69622 Villeurbanne Cedex, France}
\centerline{(e-mail~: desgrola@frcpn11.in2p3.fr)}

\medskip
\centerline{$^{(2)}$ Institute of Electron Physics,
National Academy of Sciences of Ukraine}
\centerline{ 294016 Universitetska, 21 Uzhgorod, Ukraine}
\centerline{(e-mail~: iep@iep.uzhgorod.ua)}

\medskip
\centerline{$^{(3)}$ Bogolyubov Institute for Theoretical 
Physics,
National Academy of Sciences of Ukraine}
\centerline{Metrologichna st., 14b, 252143 Kiev, Ukraine}
\centerline{(e-mail~: martynov@gluk.apc.org)}
\bigskip

\bigskip

\bigskip
{\bf Abstract}
\medskip

\hor
Alternatives for describing the nonlinear behavior of the 
first diffraction
cone in differential $pp$ and $\bar pp$ elastic 
cross-section are investigated.
High quality fits to the data are presented. We show  that 
the presence
in the Pomeron amplitude of two terms with different $t$ 
dependences is strongly
suggested by the data, hinting at a non-factorizable Pomeron
even in the field of purely hadronic reactions.
The available data, however, do no allow to choose
among a nonlinearity in the residues or in the Pomeron 
trajectory or in both.
In all cases, we find an effective slope of the trajectory 
larger
than the one currently used. A nonlinear trajectory with the 
fitted
parameters is used for predicting the mass and the width of 
the 2$^{++}$
glueball. An
excellent agreement is found with the X(1900) candidate from 
the WA91
experiment.

\vfill\eject

\vglue 1.cm
{\bf 1. Introduction}
\bigskip
\hor
The major efforts about the Pomeron problem seem to be now 
concentrated in the
Deep Inelastic Scattering (DIS) physics (see for example the 
new experiments at
HERA and their theoretical interpretations in [1]).
However, at the same time, the r\^ole
of the Pomeron in the "old" hadronic high energy physics is 
still theoretically
investigated. The subject is now revitalized by projected 
experiments,
such as the PP2PP [2] and TOTEM [3] projects at RHIC and LHC.
The reason of this two-fold effort and interest
comes mainly from the realization that low - $x$
physics is probably
strongly related to high energy low -$p_t$ physics, through 
a unique mediator:
the Pomeron,
which can be qualified as "soft" or "hard", depending on 
which aspect of this
object one wants to focuse.

\hor
This paper is concerned with the manifestations of the 
Pomeron in high energy
$pp$ and $\bar pp$ elastic scattering (see the rewiews [4a] 
and references
therein).
Many efforts (for example [4b]) have been devoted to the 
construction of the
related amplitude, able to describe the data in a wide range 
of centre-of-mass
energies ($\sqrt s$) and 4-momentum transfer (squared) ($t$) 
in the process.
However, it is well-known that, in the intermediate and large
$t$-region,  the Pomeron does not contribute alone to the 
process~: in that
kinematical region where the angular distributions exhibit a 
"dip-bump"
structure followed by a "second cone",
an interference of several terms (Pomeron, Odderon,
Reggeons....) and multiple exchanges between them enter into 
the game (mainly
multiple Pomeron and Odderon exchanges). To clarify, the 
contribution of the bare Pomeron
in the amplitude can be constructed
as an input -or Born approximation- which dominates at $t=0$ 
and at small
$|t|$, corresponding to the "first cone".
It is not excluded that the determination of the Pomeron 
contribution
from an overall $|t|$ consideration is strongly biased by 
the intermediate
and large $|t|$ data, where multiple exchanges interfere and 
the relative
contribution of the bare Pomeron is hindered.
Thus, in our opinnion, a clear signal of the Pomeron can be 
seen at high
energy and small $|t|$ and it is therefore important to 
analyse
first of all this kinematical region.

\smallskip

\hor
The data on the first cone were analyzed long ago in a wide 
energy
region.
The subsequent appearance of the first dip and second cone 
slightly shifted the
focus of the studies.
However, since this pioneering times, many precise 
measurements of the angular
distribution have been performed at small $|t|$, and it is 
necessary
to come back to the original first goal of analysing the 
elastic scattering
data to see what more we can learn from high quality fits to 
high precision
experiments.

\smallskip

\hor
Our strategy for the construction of a complete
model aiming to reproduce the available experimental data 
step by step,
in a wide domain of $\sqrt s$ and $t$ will be the following~:

1. First we select, as input Pomeron  and  preasymptotic 
$f$- and
$\omega$-Reggeons amplitudes, a simple parametrization 
fitting the $t=0$ data
and we fix the relevant parameters.

2. Next, we extend the analytical expression for scattering
amplitude to take into account the small $|t|$ -region.
From a fit to the corresponding
data, we determine the new parameters controling the 
behavior of the
amplitude in that region.

3. Finally, we incorporate
in the total amplitude $A(s,t)$  terms which are important 
at $|t|> 1$ GeV$^2$
and adjust the corresponding new parameters to the data in 
the intermediate
and high $|t|$-range, keeping unchanged the part of the 
amplitude and the
parameters determined previously.

\hor
This procedure can be repeated if necessary. We mean that, 
to perform a fine
tuning of all the parameters, giving the best fit to the 
data, it
is possible, on a given step, to allow a weak variation of 
the parameters
obtained from the previous step. In our opinion, such a 
minimization procedure
improve our understanding of the physical meaning of each 
term introduced
phenomenologically in the amplitude.

The first step has been performed in [5], where we have 
considered four
models of the Pomeron at $t=0$ and obtained the 
corresponding parameters. In
the present paper, we continue the above proposed program , 
performing the
second step, and investigating
$pp$ and $\bar pp$ elastic scattering at small $|t|$, in the 
region of the
first diffraction cone.
We come to the remarkable conclusion that a non-factorizable 
Pomeron is more
suitable for describing this region than a factorizable one.
This, while not ruled out on physical grounds came somewhat 
of a surprise since
most attempts of fitting elastic data have assumed 
factorizability of the
Pomeron. The privilege of our present investigation however 
resides  in the
high quality fits we can produce.
\smallskip

\hor
In the next section, we give theoretical and phenomenolgical 
arguments in
favor of a nonlinear Pomeron trajectory.
In the section 3, we define the scattering amplitudes and 
the set of chosen
experimental data. In the section 4, we discuss the obtained 
results of our
fit to the data, together with the properties of the Pomeron 
trajectories under
consideration, yielding predictions for the mass and the 
width of the first
lightest glueball candidate associated with the Pomeron. A 
summary of our
conclusions is presented in a last
section.

\bigskip
\bigskip

{\bf 2. Nonlinear Pomeron trajectory}
\bigskip
\hor
It is well known that the Reggeon trajectories $\alpha(t)$ 
have thresholds
related to the physical thresholds of the amplitude in the 
$t$-channel
[6]. Above a threshold, the trajectory has an imaginary part
related to the width $\Gamma_R$ of a resonance peak at the 
point $t_R$ where
the real part of trajectory equals the spin $S$ of the 
resonance
$$\Re{\hbox{e}}\, \alpha(t_R)=S \, 
.                              \eqno(1)$$
Then, in accordance with the Breit-Wigner formalism,
$$\Gamma_R = {\Im{\hbox{m}}\, \alpha(t_R)\over
{\Re{\hbox{e}}\, \alpha'(t_R)\, M_R}}\ ,\quad M_R\, =\, 
\sqrt{t_R}\, ,\eqno(2)$$
where $\alpha'(t_R)$ is here the derivative taken at the 
resonance and
$M_R$ is the "mass" associated to the resonance .
It is necessary to emphasize that such a trajectory cannot 
be a linear function
of $t$. Moreover there are severall arguments [7] to think 
that trajectory
$\alpha (t)$ cannot linearly increase with $t$, {\it i.e.}
$$ |\alpha (t)/t|\rightarrow 0 \qquad\hbox{when} \quad 
|t|\rightarrow \infty.\eqno(3) $$

This is certainly valid for the trajectories with 
resonances. As for the
Pomeron, up to now it is not clear whether resonances
exist or not. From QCD arguments as well as in analogy with 
Reggeons, one
expects that observable particles
(glueballs) should be found on the Pomeron trajectory for 
integer spins larger
than one. Candidates for this role of glueballs are seen
in the experiments [8], but the question is still open (some 
claiming
experimental evidence, others claiming the absence of clear 
experimental
signal). Nevertheless, if such resonances
exist, the Pomeron trajectory should be nonlinear. It has
thresholds and the corresponding minimal $t$-value is 
related to the lightest
mass of
hadronic state with vacuum quantum numbers. This is a two 
pions state. So the
Pomeron trajectory should have a threshold at 
$t_P=4m_{\pi}^2$.

\hor
Additional
arguments in favor a nonlinear trajectory (for $t$ not close 
to zero) can be
found also in [9]. In that paper, it has been shown that the 
unitarity
inequality
$$\Im\hbox{m}\,  H(s,b)>0                                
\eqno(4)$$
for an elastic scattering amplitude $H(s,b)$ in the 
impact-parameter ($b$)
representation as well as the correct asymptotic behavior of 
this amplitude
$$H(s,b)\sim \exp
(-b/b_0),\quad b_0=\hbox{constant}\ ,\quad
\hbox{when}\ b\to \infty                                    
\eqno(5)$$
cannot be satisfied, even after an eikonalization has been 
performed, if the
input Pomeron has a linear trajectory.

\hor
Finally, we must
note that a nonlinearity of the Pomeron trajectory $\alpha_P 
(t)$
is to be reflected into
the curvature of differential cross-sections (${d\sigma\over 
dt}$) at very high
energy ($\sqrt s$) and small squared transfer ($|t|)$. For 
the linear
case ($\alpha_P (t)\, = \, \alpha_P (0)+\alpha'_P t$),
the local slope $B(s,t)$ of ${d\sigma\over dt}$ is 
$t$-independent
and of course the local curvature $C(s,t)$, related to the 
second derivative, is
zero.
$$B(s,t)={\partial\over\partial 
t}\left(\ln{d\sigma(s,t)\over dt}\right)
\, \simeq \,2\, \alpha'_P\ln{s \over s_0}\, , \qquad
C(s,t)={1\over 2}\left({\partial\over\partial 
t}B(s,t)\right)\, =0\, ,\eqno(6)$$
while for nonlinear trajectories, one obtains for the slope 
and the curvature
at very high energy
$$B(s,t)\, \simeq\,2\, {d\alpha_P(t)\over dt}\ln{s \over 
s_0}\, , \qquad
C(s,t)\, \simeq\,  {d^2\alpha_P(t)\over dt^2}\ln{s \over 
s_0}\, . \eqno(7)$$

If the rising of a nonlinear $\alpha_P(t)$ with $|t|$ is 
slower than $t$,
then the second derivative is positive {\it i.e} at a
sufficiently large $s$, the curvature will be positive.
However, at the available energies, below $\sqrt s= 1.8$ TeV,
$C$ is found positive  and at $\sqrt s= 1.8$ TeV, $C$ is 
roughly zero [10].
Thus in accordance with [10] the Tevatron energy is close to the 
transition between the
preasymptotic ($C>0$) and asymptotic ($C<0$) behaviors of
${d\sigma(s,t)\over dt}$. Our model predicts another behavior of $C$; it 
stills positive and rising with energy. Future experiments, we hope, can 
help to select the realistic model. 

\hor
On the other hand, given that the elastic scattering 
amplitude does not have a
singularity at $t=0$, the Pomeron
trajectory also must have a regular behavior at $t=0$, if 
the Pomeron is an
isolated singularity in the $j-$plane. This is not the case 
for models in which
the Pomeron is a pair of moving cuts colliding at $t=0$ (for 
example, in an
eikonalized model with as input a simple pole having an 
intercept larger than
1). We note that
if the Pomeron trajectory is assumed to behave linearly
($\alpha_P (t)\approx \alpha_P (0)+\alpha'_P t$) when 
$t\approx 0$,
the Pomeron pole cannot be harder
than a double one. In fact, let the elastic partial 
amplitude have the pole
with arbitrary hardness $\mu +1$
$$a(j,t)\sim {1\over (j-1-\alpha'_Pt)^{\mu+1}}\ 
.                  \eqno(8)$$
One can easily obtain
 the following asymptotic behaviors of the total and
elastic cross-sections~:
$$\sigma_{tot}(s)\sim \ln^{\mu}(s/s_0), \quad 
\sigma_{el}(s)\sim
\ln^{2\mu-1}(s/s_0), \quad s_0=1\hbox{ GeV}^2\ .\eqno(9)$$
Because $\sigma_{el}<\sigma_{tot}$, we must have $\mu\le 1.$
In other words, if the Pomeron trajectory is linear for $t$ 
close to zero,
then the Pomeron cannot have a higher singularity than a 
double $j-$pole.

\hor
Thus, we prefer to limit our present analysis and 
description of the available
experimental data to a double Pomeron pole
($\mu=1$)  as well as to a simple Pomeron pole ($\mu=0$) and
use a nonlinear Pomeron trajectory. In addition, we consider 
also the
standard linear case in order to
compare different possibilities and see the effects of 
nonlinearity
of the trajectory and of the residue functions (see below).
The choice of a common set of experimental data is crucial 
for discussing the
meaning of this comparison.

\bigskip

{\bf 3. Definition of the amplitude and choice of 
experimental data}
\bigskip

{\it {Models for the scattering amplitude}}
\medskip

\hor
As usual in the Regge approach, we write the following form 
for the
$\bar pp$ and $pp$ elastic scattering amplitudes
$$A^{\bar pp}_{pp}(s,t) = 
P(s,t)+f(s,t)\pm\omega(s,t)+C^{\bar pp}_{pp}(s,t)\, ,\eqno(10)$$
where $P(s,t)$, $f(s,t)$, $\omega(s,t)$ are respectively the 
Pomeron,
the $f$- and the $\omega$-Reggeon contributions; $C^{\bar 
pp}_{pp}(s,t)$
is the standard Coulomb amplitude, which has been calculated 
according to the
procedure by West and Yennie [11].
We use the following normalization for the dimensionless 
amplitude
$$\sigma_{tot}={4\pi\over s}\Im\hbox{m}\,A(s,t), \qquad
{d\sigma\over dt}={\pi\over s^2}|A(s,t)|^2\,. \eqno(11)$$

\hor
The $f$-Reggeon has a standard parametrization
$$f(s,t)=g_f{\tilde s}^{\alpha_f(t)}\; e^{b_ft}\ ,\hbox{ 
with}\
\alpha_f(t)=\alpha_f(0)+\alpha_f' t\, .\eqno(12)$$
Here and in what follows $\tilde s=-is/s_0$ and $s_0=$ 1 
GeV$^2$.
In the $\omega$-Reggeon contribution
$$\omega(s,t)=i\, g_{\omega}\left[1+t(z_1+z_2\ln\tilde 
s)\right]
{\tilde s}^{\alpha_{\omega}(t)}\; e^{b_{\omega}t}\ ,\hbox{ 
with}\
\alpha_\omega (t)=\alpha_\omega (0)+\alpha'_\omega t\, .\eqno(13)$$
we insert an additional factor $1+t(z_1+z_2\ln\tilde s)$ which
phenomenologically describes the cross-over effect, {\it 
i.e.} the crossing of
the $\bar pp$ and $pp$
differential cross-sections at some transfer 
$t=t_{\omega}<0$. This is
illustrated in  Fig.1 for $\sqrt{s}=13$ GeV$^2$. Comparing the
different possibilities to describe this phenomenon, we 
observe that the
best description of the data is obtained when $t_{\omega}$ 
is moving down to
zero when the energy increases. This is a fine effect, 
however ignoring it
leads to a higher value of the $\chi^2$ (see the numerical 
results in the next
section).

\hor
Choosing the Pomeron contribution is a more delicate and 
complicated
subject. One of our aims in the present study is to 
investigate the effects of
nonlinearity of the Pomeron trajectory. However, the 
nonlinear $t$- behavior
of $\ln {d\sigma\over dt}$  can eventually be described by 
residue functions
$F^2(t)$ which can differ from usual ones $\exp(b_0t)$ with 
$b_0$ constant.
For that reason, we study a more general parametrization
$F^2(t)=\exp\phi(t)$ (see below).

\hor
We consider two types of Pomeron singularity, first a double 
$j$-pole with
trajectory of unit intercept; we call it "model A"~:
$$\hbox{\bf model  A}\quad\qquad\qquad \alpha_P(0)=1\, 
,\quad\quad
P(s,t)=\left[g_0F_0^2(t)+g_1F_1^2(t)\ln{\tilde s}\right]
{\tilde s}^{\alpha_P(t)} \,,\eqno (14)$$
where, in order to limit the free parameters, we choose the 
residue functions
$$ F_1^2(t)=\exp{(\beta t)}\, F_0^2(t)\,,\eqno(15)$$
$$F_0^2(t)=\exp(\phi(t)), \qquad \phi(0)=0.\eqno(16)$$
Our second choice will be a simple $j$-pole with an 
intercept larger than 1;
we call this model of supercritical Pomeron "model B"~:
$$\hbox{\bf model  B}\quad\quad \alpha_P(0)=1+\Delta>1\, 
,\quad
P(s,t)=\left[g_0F_0^2(t)  {\tilde s}^{-\Delta} 
+g_1F_1^2(t)\right]
{\tilde 
s}^{\alpha_P(t)},\eqno (17)$$
with the same simplificative choice for the residue 
functions as in the
preceeding case.
Here, we would like to note that both expressions (14) and 
(17) for the Pomeron
contribution cannot satisfy the factorization condition (if 
$\beta\ne 0$).
This is due to the preasymptotic terms $g_0 F_0^2 \tilde  
s^\alpha$.
For "model A" for example, the first and second terms 
correspond to the case of
a single and double $j$-pole respectivly. Of course, the 
double pole
contribution taken alone (as the leading term at high 
energy) would satisfy
the factorization condition.
Thus, the condition $\beta\ne 0$ can be considered as an 
indication of the
nonfactorizabiliy of the Pomeron at present energies
(and this will, indeed, be one of the conclusions following 
our analysis,
see below).
For both models of the Pomeron, we consider and compare a 
few possibilities
(see also [12])~:

{\it (i)}
$$\alpha_P(t) = \alpha_P(0)+\alpha'_P 
t+\gamma_P\left(1-\sqrt{1-t/t_P}\right), \eqno(18)$$
$$\phi(t) = b_0 t+\gamma_0\left(1-\sqrt{1-t/\tau_0}\right)\,\eqno(19)$$
{\it (ii)}
$$\alpha_P(t) = \alpha_P(0)+\alpha'_P t+
\gamma_P\left(1-(1-t_P/t)\ln(1-t/t_P)\right)\,. \eqno(20)$$
$$\phi(t) = 
b_0t+\gamma_0\left(1-(1-\tau_0/t)\ln(1-t/\tau_0)\right),\eqno(
21)$$
{\it (iii)}
$$\alpha_P(t) = \alpha_P(0)+\alpha_P't-\gamma_P
\ln{1+\sqrt{1-t/t_P}\over 2}\, ,\eqno(22)$$
$$\phi(t) =
b_0t-\gamma_0\ln{1+\sqrt{1-t/\tau_0}\over 2} \, , \eqno(23)$$
where in accordance with arguments given in the previous 
section
$t_P=4m_{\pi}^2.$

The following remarks are in order~:

\hor -
For simplicity, the nonlinear corrections have the same form 
in the
trajectories and in the function $\phi(t)$ defining the 
common factor in the
residues.

\hor -
In all cases, we write the linear term only as an
effective contribution of $N$ other nonlinear terms, related 
with the next
thresholds, because we consider a very limited region of 
small $|t|$.
More generally, for example, in case {\it (i)}, we could 
have written instead of
(18)
$$\alpha_P(t)=\alpha_P(0)+\sum_{k=1}^{N}\gamma_{P,k}(1-\sqrt{1
-t/t_{P,k}})\, ,\qquad  4m^2_\pi = t_{P,1}<t_{P,2}<...., \eqno(24)$$
so that the effective slope for the linear part of 
trajectory (8) would be
$$\widetilde{\alpha'_P}'\, =\, 
\sum_{k=2}^{N}{\gamma_{P,k}\over 2t_{P,k}}.\eqno(25)$$
\hor -
The slope of nonlinear Pomeron trajectories depends on t; in 
particular at
the origin
$$\widetilde{\alpha'_P}=\left[{d\alpha_P(t)\over 
dt}\right]_{t=0}=
\alpha'_P+{\gamma_P\over 2\nu t_P}\,\eqno(26)$$
where $\nu=1$ for (18) or (20) and $\nu=2$ for (22).

\hor -
The trajectory {\it (ii)} does not have an unwanted singular 
behavior at
$t=t_P$ like
the simplest logarithmic trajectory 
$\alpha(t)=\alpha(0)-\gamma\ln(1-t/t_P)$
(the pole of the factor $1-t_P/t$ being cancelled by the 
zero of the logarithm).

\hor -
For the purpose of comparison we have also considered the 
simple case of linear
functions
$$\alpha_P(t)=\alpha_P(0)+\alpha'_P t\,, \eqno(27) $$
$$\phi(t) = b_0 t\,.\eqno(28)$$
\bigskip

{\it {Experimental data}}
\medskip

\hor
In order to determine the parameters  which control the 
$s$-dependence of
$A(s,0)$, we use practically the same set of data on the 
total cross-sections
$\sigma_{tot}(s)$ and on the ratios
$\rho(s)=\Re\hbox{e}\,A(s,o)/\Im\hbox{m}\,A(s,0)$ as in [5] 
for $\sqrt{s}\ge 5$
GeV. The recent value of $\sigma_{tot}^{\bar p p}$ at the 
Tevatron energy [13]
is added; however, the ancient result $\rho_{\bar 
pp}=0.24\pm 0.04$ at 546 GeV
has been eliminated.
A total of 208 data has been included for $t=0$.

\hor
For the differential cross-sections ($d\sigma/dt$)
we have selected the data for energies
$\sqrt{s}>9$ GeV. However, the angular distributions at the 
Tevatron have not
been taken into account in the fit. The squared 4-momentum 
has been limited by
$|t|\le
0.5$ GeV$^2$, because at larger $|t|$ the influence of the 
dip-region becomes
visible (in particular, it can be seen quite clearly at the 
Collider energy
$\sqrt{s}=546$ GeV). To be more precise, the $t$-limit of 
the first diffraction
cone changes weakly with energy. We
think that the choosen region $|t|\le 0.5$  GeV$^2$ is 
certainly the region of a
first cone for the energies investigated here, while at 
smaller energies it
can be extended.
Several experiments, in which $d\sigma/dt$ is measured
at small $|t|$, have been reported. In the present 
analysis,  we keep only those
for which the data cover the largest range of energies and 
(or) those for which
data on both $pp$ and $\bar p p$ are available.
We include also the data for
$d\sigma/dt$ in the Coulomb-nuclear interference region, 
when they exist.
A grand total of 1288 data have been used in the overall fit.

\bigskip
                      \vfill\eject
{\bf 4. Results and discussion}
\medskip

{\bf Data at $t=0$.}

\hor
The parameters from the fit to the data at $t=0$ are given 
in Table 1 for both
versions of the Pomeron model. We notice that the double 
pole Pomeron
with $\alpha_P(0)=1$ (model "A")
gives a marginally better $\chi^2$  than the supercritical 
Pomeron with
$\alpha_P(0)  > 1$ (model "B").

\hor
The theoretical total cross-sections
$\sigma_{tot}(s)$ and ratios $\rho(s)$ are compared to the 
data for $pp$ and
$\bar pp$ elastic scattering respectively in Fig.2 and Fig.3 
for the
double pole (model "A") only.
The results for both versions would be undistinguishable on 
these figures except at
the Tevatron energy. At such an energy, and consequently at 
higher energies,
the supercritical model predicts a total cross-section 
greater than the double
pole model does (for example at the 14 TeV of the LHC~: 110 
mb instead of 102
mb).

\medskip

{\bf Data at $|t|\le 0.5$ GeV$^2$}.

\hor
In accordance with the strategy laid down in the 
introduction, all parameters
of Table 1, are fixed when fitting the amplitude (with its 
new parameters) to
the $t\ne 0$ data.
The three nonlinear parametrizations ({\it (i,ii,iii)})
have been tested in both models "A" and "B" and compared to 
the linear ones.

\hor
As an example of our results, we show in Fig.4 the 
comparison of the
theoretical angular distributions to the data in the first 
cone, in
the case of model "A" with non linear component either in 
trajectory and-or-
in residue giving a high quality fit
(with $\chi^2_{d.o.f} \simeq$ 1.8).  A special
attention is paid in Fig.5 where the very small $t$ region 
of the
nuclear-Coulomb interference is exhibited. The relevant 
parameters are listed
in Table 2.

First, we compare the models "A" and "B" of the Pomeron and 
comment the
contributions of the Reggeons.
Then, we discuss and interpret the results obtained in terms 
of
the residue functions and the trajectories for the Pomeron.

1. {\bf Quality of the fits.}

In all variants ({\it i - iii}) considered, the quality of
description of data is slightly better for the model "A" of 
the Pomeron
than for the model "B" (we draw such a conclusion from the 
$\chi^2_{d.o.f}$,
but the plots are sometimes undistinguishable by eye).
We failed to improve the model "B"
by choosing residue functions of the "QCD" form $(1-{t\over 
t_i})^{-4}$.

\smallskip
                               \vfill\eject
2. {\bf Cross-over.}

Account of the cross-over factor in the $\omega$-part of the
amplitude (13) allows to improve the description of data.
In fact, setting 1
instead of the factor $1+t(z_1+z_2\ln\tilde s)$ and 
refitting the remaining
free parameters increases the value of $\chi_{d.o.f}^2$ by
30\% in all versions under consideration.
In addition we find the value of $b_\omega $ to be quite 
relevant and we fix it
equal to zero.
\smallskip

3. {\bf Secondary reggeons.}

Owing to the fact that the low-energy angular distributions 
(down to
9 GeV, where the $f$- and $\omega$-Reggeons are very 
important) enter in the
fit,
we were able to adjust the slopes of the secondary 
trajectories
$\alpha'_f$ and $\alpha'_\omega$. The values we obtained are 
always noticeably
larger than those commonly reported. The same observation 
has been made when
eikonalizing amplitudes to describe ${d\sigma\over dt}$ 
beyond the first cone
[14]. Maybe it is suitable to use nonlinear parametrization 
for $f$- and
$\omega$ trajectories as we do for Pomeron.
\smallskip

4. {\bf Nonfactorizability of the  Pomeron.}

A good description of the data for
${d\sigma\over dt}$ is obtained only if $F_0\ne F_1$ (or 
$\beta\ne 0$) in (15).
As already noted in Sect.3, this is  an evidence of 
nonfactorizability of the
Pomeron at available energies. Since $\beta<0$, 
the deviation from
factorizability is decreasing with $|t|$. We emphasize that we 
are concerned with
the low $|t|$ region and simple Pomeron alone (without 
rescatterings or cuts).
At truly asymptotic eneergies, the factorization is restored 
(the leading
term is with $F_1^2(t)$ in (14),(17)).
The observed effect of nonfactorization we claim has nothing 
to do
with subleading contributions ($f,\omega,$... {\it etc.}) 
which are
explicitly taken into account. Notice only that, contrary to 
DIS analyses, our
conclusion rests on the extremely good quality of both data 
and fits.
\smallskip

5. {\bf Non linear effects.}

An explicit form of the function $\phi(t)$ appears to be 
largely irrelevant
(we believe this is due to the small $t$-domain investigated 
here;
choosing $\phi_i(t)$ will be more crucial for large $|t|$
, as well as choosing an exponential form (16) to
define the residue functions themselves). However, we found 
that the free
parameter $\tau_0$
always tends to its smallest possible value 
$\tau_0=4m_\pi^2$  value and
thus we can fix it at this limit.

We observed that in the versions with a non linear function 
$\phi(t)$, the
$\gamma_P$ parameter tends to zero, {\it i.e.}
the Pomeron trajectory tends to be a linear one.
However, for versions with linear $\alpha_P(t)$ and non 
linear $\phi(t)$
and versions with nonlinear $\alpha_P(t)$ and linear 
$\phi(t)$ the difference in
$\chi^2$ is less that 2\%. Only when $\alpha_P(t)$ and 
$\phi(t)$
are both linear, the $\chi^2$ increases by 20\%.

Therefore, in accordance with the theoretical arguments 
discussed in Sect.2, in
favor of a nonlinear trajectory, we prefer the versions with
nonlinear $\alpha_P(t)$ and linear $\phi(t)$.

\smallskip

6. {\bf Slope of the Pomeron trajectory.}

Another important and interesting result of our 
investigation is that we  find
quite a larger slope of the Pomeron trajectory  than the
"world value" $\alpha'_P=0.25$ GeV$^{-2}$.
In the linear cases the results of fits gives even larger 
values.
For the non linear cases, as already said, the slope is 
$t$-dependent.
The calculation of ${d\alpha_P(t)\over dt}$ at the frontiers
(for $t=$0 see (26)) and at the center of the first cone are 
shown in
Table 3, for the same choices of parametrization as above in 
Table 2.
The trend with $t$ is in accord with earlier investigations 
[15].
\smallskip

\hor
The conclusions of the previous points are valid for both 
models "A" and "B".
\smallskip

7. {\bf Pomeron trajectory and 2$^{++}$ glueball.}

We now present the results concerning the prediction of a 
2$^{++}$ glueball
obtained in the dipole Pomeron model with unit intercept 
("A").
We consider the non linear trajectories with a threshold at
$t_P=4m_\pi^2$. These trajectories have an imaginary part at 
$t>t_P$. The
behavior of the real and imaginary parts are shown in Fig.6 
in the particular
case of the three Pomeron trajectories  listed in Table 2.
We found that in all variants giving a similar $\chi^2$, the 
Chew-Frautschi
plots show also similar predictions for the mass and the 
width of the resonance
 $$ 1.89 \, <\, M_g (\hbox{GeV})\, <\, 1.92, $$
$$ 100\,<\, \Gamma_g (\hbox{MeV})\, <\, 400. $$

One notes quite a good agreement with the result of the WA91 
experiment [8b].
The measured values for the X(1900) which could be a single 
state with
I(J$^{PC}$) = 0(2$^{++}$) are $$M_g=1.926\hbox{ GeV} \pm 
12\hbox{ MeV}, $$
$$\Gamma_g=(370 \pm 70)\hbox{ MeV}.$$

\hor
Note that, among the various case of nonlinear trajectories 
examined here,
the  {\it (i)} case
(18) involving a square root leads to the width in best 
agreement with the
experimental value.

\hor
A larger glueball mass is predicted in [16], where the 
nonlinear Pomeron
trajectories with an intercept larger than 1 is used to 
describe the data on
the whole $t$-domain for which dat exist.
\bigskip

{\bf 5. Conclusions}
\medskip

\hor
To summarize, we emphasize once more that the
nonfactorizable form of the Pomeron amplitude as well as the 
nonlinearity
of its trajectory $\alpha_P(t)$ or/and of the
function $\phi (t)$ entering in its residue is strongly 
suggested by the data.
To clarify the question whether the nonlinear behavior of 
the first diffraction
cone is due to a
nonlinear trajectory or to complicated residue functions, it 
would be
necessary to have more precise data at high energies. The 
projected
measurements at RHIC and LHC energies (where the $f$-Reggeon 
contribution is
expected to be negligible) will certainly bring useful 
informations
and, possibly, a reliable answer.

\hor
An important result of this work is a larger than usually 
used effective slope
of the Pomeron
trajectory we encounter in all successful parametrizations. 
It decreases with
$|t|$ but for the first cone it varies within the limits:
$$0.32\ < {d\alpha(t)\over dt}\ (\hbox{ GeV}^{-2}) < 0.46\, ,
\qquad \hbox{for}\quad - 0.5 \hbox{ GeV}^{2} \, < \, t \, 
<0.$$

\hor
Finally, we point out that the trajectory parameters are 
found very close to
each others
in the various parametrizations. They yield a mass and width 
of the first
2$^{++}$ glueball very close to the X(1900) observed by the 
WA91
collaboration. In our opinion, one may interpret this as a 
confirmation of a
possible determination of the true Pomeron from the study of 
small $|t|$ data
on the elastic hadron scattering.

\bigskip
{\bf Acknowlegments}
We would like to thank E. Predazzi for reading the 
manuscript and enlightning
comments on the factorization problem.
M. Giffon is thanked for useful discussions.
Two of us (E.S.M and A.I.L) thank the I.P.N at Lyon for kind 
hospitality;
E.S.M is supported by IN2P3, A.I.L by I.P.N.L.

\vfill\eject

{\centerline{\bf References}}
\bigskip

\item{[1]} M. Haguenauer (editor)
Proceedings of the 6$^{\rm th}$ International Conference on 
Elastic and
Diffractive Scattering, Blois, France (6$^{\rm th}$ "Blois 
workshop", June
1995), {\it to be published} (Editions Frontires).
\smallskip

\item{[2]} W. Guryn and R7 Coll. {\it The "PP2PP" project.
Proposal to measure total and elastic $pp$ cross sections at 
RHIC},
6$^{\rm th}$ International Conference on Elastic and
Diffractive Scattering, Blois, France (6$^{\rm th}$ "Blois 
workshop", June
1995), {\it to be published}.
\smallskip

\item{[3]} M. Buenerd et al. {\it The TOTEM project at LHC},
6$^{\rm th}$ International Conference on Elastic and
Diffractive Scattering, Blois, France (6$^{\rm th}$ "Blois 
workshop", June
1995), {\it to be published} and references therein.
\smallskip

\item{[4]} a) G. Matthiae.
Rep. Prog. Phys. {\bf 57} (1994) 743~;

M. Bertini, M. Giffon.
Phys. Part. Nucl. {\bf 26}-1 (1995) 12~;

M. Bertini {\it et al.} La Rivista del Nuovo Cimento {\bf 
19} (1996) 1.

b) C. Bourelly, J. Soffer, T.T. Wu.
Phys. Rev. {\bf D 19} (1979) 3249; {\it ibid} Nucl. Phys. 
{\bf B 247} (1984)
15.~;

P. Gauron, B. Nicolescu, E. Leader.
Nucl. Phys. {\bf B 299} (1988) 640~;

A. V. Wall, L.L. Jenkovszky, B.V. Struminsky.
Sov. Jour. Part. Nucl. {\bf 19}-1 (1988) 77~;

P. Desgrolard, M. Giffon, E.Predazzi.
Zeit. Phys. {\bf C 63} (1994) 241~;

\smallskip

\item{[5]} P. Desgrolard, M. Giffon, A. Lengyel, E. Martynov.
Nuovo Cim. {\bf 107 A} (1994) 637.
\smallskip

\item{[6]} P.D.B. Collins. {\it An introduction to Regge 
theory and high-energy
physics}, University Press, Cambridge (1977).
 \smallskip

\item{[7]} R.W. Childres. Phys Rev. {\bf D 2} (1970) 1178~;

E. Predazzi, H. Fleming. Nuovo Cim. Lett. {\bf 4} (1970) 556~;

H. Fleming. Nuovo Cim. {\bf 14 A} (1973) 215~;

A.A. Trushevsky {\it "Asymptotic behavior of Boson Regge 
trajectories"}
Preprint ITP-75-81 E, Kiev,1975.

\smallskip

\item{[8]} a) see for example Review of Particle Physics. 
Phys. Rev. {\bf D 54}
(1961) 1 and references therein;

b) WA91 Collaboration, S. Abatzis  {\it et al.}. Phys.Lett. 
{\bf B 324} (1994)
 509.

\smallskip

\item{[9]} A.V. Akkelin, E. S. Martynov.
Sov. J. Nucl. Phys. {\bf 55} (1992) 1555.
\smallskip

\item{[10]} M. Block, K. Kang and A.R. White.
Intern. Jour. Mod. Phys. {\bf 7} (1992) 4449.
\smallskip

\item{[11]} G.B. West, D.R. Yennie. Phys., Rev. {\bf 172} 
(1968) 1413.
\smallskip

\item{[12]} Various parametrizations for Pomeron nonlinear 
trajectories are
also considered for description of angular distribution at 
large transfer;
relevant references can be found in L. Jenkovszky, La 
Rivista del Nuovo Cimento
(1987) 1.
\smallskip

\item{\bf [13]} CDF Collaboration, F. Abe {\it et al.}
Phys. Rev. {\bf D 50} (1994) 5550.
\smallskip

\item{\bf [14]} M. Giffon, private communication.
\smallskip

\item{\bf [15]} M. Giffon, R. Nahabetian, E. Predazzi, see 
for instance Phys.
Lett. {\bf B 205} (1988) 363, where earlier references can 
be found.
\smallskip

\item{\bf [16]} P. Desgrolard, L.L. Jenkovszky, A.I. Lengyel.
{\it in} "Strong interaction at long distances", Edited by 
L.L. Jenkovszky,
Hadronic Press Inc., Palm Harbor (1995) 235.

\vfill\eject

\def\init{\tabskip 0pt\offinterlineskip}
\def\crr{\cr\noalign{\hrule}}

$$\vbox{\init\halign to 12.cm{
\strut#&\vrule#\tabskip=1em plus 2em&
\hfil$#$\hfil&
\vrule#&
\hfil$#$\hfil&
\vrule#&
\hfil$#$\hfil&
\vrule#\tabskip 0pt\crr
&&                  &&{\bf Model\ A }&&{\bf Model\ B } &\crr
&&\chi^2_{d.o.f}    &&   1.13             &&   
1.15             &\crr
&&g_f               &&  -24.5            &&  
-15.8             &\cr
&&g_\omega          &&   9.16           &&    
9.29            &\cr
&&\alpha_f(0)       &&   0.808            &&    
0.714           &\cr
&&\alpha_\omega(0)  &&   0.451            &&    
0.445           &\cr
&&g_0               &&   7.98             &&   
6.68            &\cr
&&g_1               &&  -1.48             &&    
-8.59           &\cr
&&\Delta            &&  0            &&    0.064           
&\crr
}}$$

\centerline {\bf Table 1 }
\medskip

Parameters of the scattering amplitude for the $f$-Reggeon 
(12),
$\omega$-Reggeon (13) for versions "A" (double pole, (14)) 
and "B"
(supercritical, (17)) of the Pomeron, fitted to the 
$t=0$-data.

\bigskip

$$\vbox{\init\halign to 12.cm{
\strut#&\vrule#\tabskip=1em plus 2em&
\hfil$#$\hfil&
\vrule#&
\hfil$#$\hfil&
\vrule#&
\hfil$#$\hfil&
\vrule#&
\hfil$#$\hfil&
\vrule#\tabskip 0pt\crr
&&          
\hbox{Parameters}&&\hbox{{\it(i)}}&&\hbox{{\it(ii)}}&&
\hbox{{\it(iii)}}&\crr
&&\alpha'_f\ (\hbox{ GeV}^{-2})  &&  1.27  && 1.29 && 
1.25         &\cr
&&b_f\  (\hbox{ GeV}^{-2})       && 0.443  && 0.387 && 
0.499    &\cr
&&\alpha'_\omega\  (\hbox{ GeV}^{-2})&&   0.980  && 0.975 
&&0.967     &\cr
&&b_\omega\ \hbox{(fixed)} &&   0.  &&0.  &&   0.  &\cr
&&z_1   \ (  \hbox{ GeV}^{-2})       &&  -6.01   &&-6.22  
&&-5.77        &\cr
&&z_2   \ (   \hbox{ GeV}^{-2})      &&   2.62   &&2.68  && 
2.55    &\cr
&&\alpha'_P\ ( \hbox{ GeV}^{-2})     &&  0.256   &&0.288  && 
0.294   &\cr
&&\gamma_P               &&   0.027  &&0.022  && 0.044  &\cr
&&\beta         \ (  \hbox{ GeV}^{-2})&& -2.23 && -2.33  && 
-2.21      &\cr
&&b_0     \ (  \hbox{ GeV}^{-2}) &&    5.32  && 5.47  &&  
 5.32 &\cr
&&\gamma_0 \hbox{(fixed)}              &&   0.  &&0.  && 0.  
&\crr
}}$$

\centerline {\bf Table 2 }
\smallskip

Parameters of the scattering amplitude for the $f$-Reggeon 
(12),
the $\omega$-Reggeon (13) for version "A" of the Pomeron 
(double pole
model "A", in the three cases of non linear trajectory 
(18)(20)(22) and linear
residue (28))
fitted to the $|t|\le 0.5$ GeV$^2 $-data.

For each case ({\it (i),(ii), (iii)}), the two other
possibilities involving a nonlinear component in the 
trajectory and/or
in the residue ($\gamma_P $ and $\gamma_0\ \ne 0$,$\gamma_P\ 
=0 $ and
$\gamma_0\ \ne 0$) give a same $\chi^2$
within a few percents and are undistinguishable on curves.

\bigskip
\bigskip

$$\vbox{\init\halign to 13.5 cm{
\strut#&\vrule#\tabskip=1em plus 2em&
\hfil$#$\hfil&
\vrule#&
\hfil$#$\hfil&
\vrule#&
\hfil$#$\hfil&
\vrule#&
\hfil$#$\hfil&
\vrule#\tabskip 0pt\crr
&& \hbox{Eq.}&&
\left[{d\alpha_P(t)\over dt}\right]&&
\left[{d\alpha_P(t)\over dt}\right]&&
\left[{d\alpha_P(t)\over dt}\right] &\cr
&& &&{t=0}&&{t=-0.25\hbox{GeV}^2}&&{t=-0.50\hbox{GeV}^2}&\crr
&& \hbox {(18)} &&0.458&&0.358 && 0.335      &\cr
&& \hbox {(20)} &&0.438&& 0.339 && 0.320     &\cr
&& \hbox {(22)} &&0.438&&0.338&& 0.321       &\crr
}}$$

\bigskip

\centerline {\bf Table 3 }
\medskip

Effective slope of the Pomeron trajectories (in GeV$^{-2}$).
Only model "A"  parameters of Table 2 are used at the center 
and the limits of
the first cone.

                    \vfill\eject

\centerline{\bf Figures captions}
\bigskip
\bigskip

{\bf Fig.1} Cross-over of the experimental $ pp$ (triangles) 
and $ \bar
pp$ (circles) angular distributions at 13 GeV (dashed and 
solid lines
respectively are interpolations of the data).
When the energy increases, the $|t|$-value of the
crossing point goes smoothly below .15 GeV$^2$ (see the text).
\medskip

{\bf Fig.2} Fits of the total cross-sections up to the 
Tevatron energy
as calculated with version "A" for the Pomeron (see the text 
and Table 1).
\medskip

{\bf Fig.3}
Fits of the ratios of the real to the imaginary part of the
forward elastic amplitude as calculated with version "A" for 
the Pomeron
(see the text and Table 1).
\medskip
\medskip

{\bf Fig.4} Angular distributions limited to the first cone
(a factor of 10$^{-2}$ between each succesive curve is used).
The Tevatron data  ($\bar pp$ at 1800 GeV) are not included 
in the fit.
The solid curves are calculated with model "A" and include a 
non linear
component for the Pomeron either in trajectory and-or- in 
residue
(see Table 2 for the parameters).
\medskip

{\bf Fig.5} Enlargement of fig.4 showing the (very small 
$|t|$) nuclear-Coulomb
interference region.
\medskip

{\bf Fig.6} Real and imaginary parts of the trajectories 
versus the
four-momentum transfer
for the various options of the Pomeron dipole with $\delta 
=0$ (model "A")
discussed in the text. The solid lines correspond to (i); 
dashed lines to (ii);
dotted lines to (iii). All these Chew-Frautschi plots
predict a mass and a total decay width of the first 
candidate glueball with
$J^{PC}=2^{++}$ in agreement with the measurements of the 
X(1900) by  Wa91
Collab. [8b] (slight differences would occur in predictingg 
the next
recurrence ($J=4$), but the predicted order of magnitude is 
the same).

The lower part of the figure exhibits the non linearity of the real part 
of the trajectory at low $t$ (over a range symmetrical to the first cone).

\input epsf
\epsfxsize=16 true cm
\centerline{\epsfbox{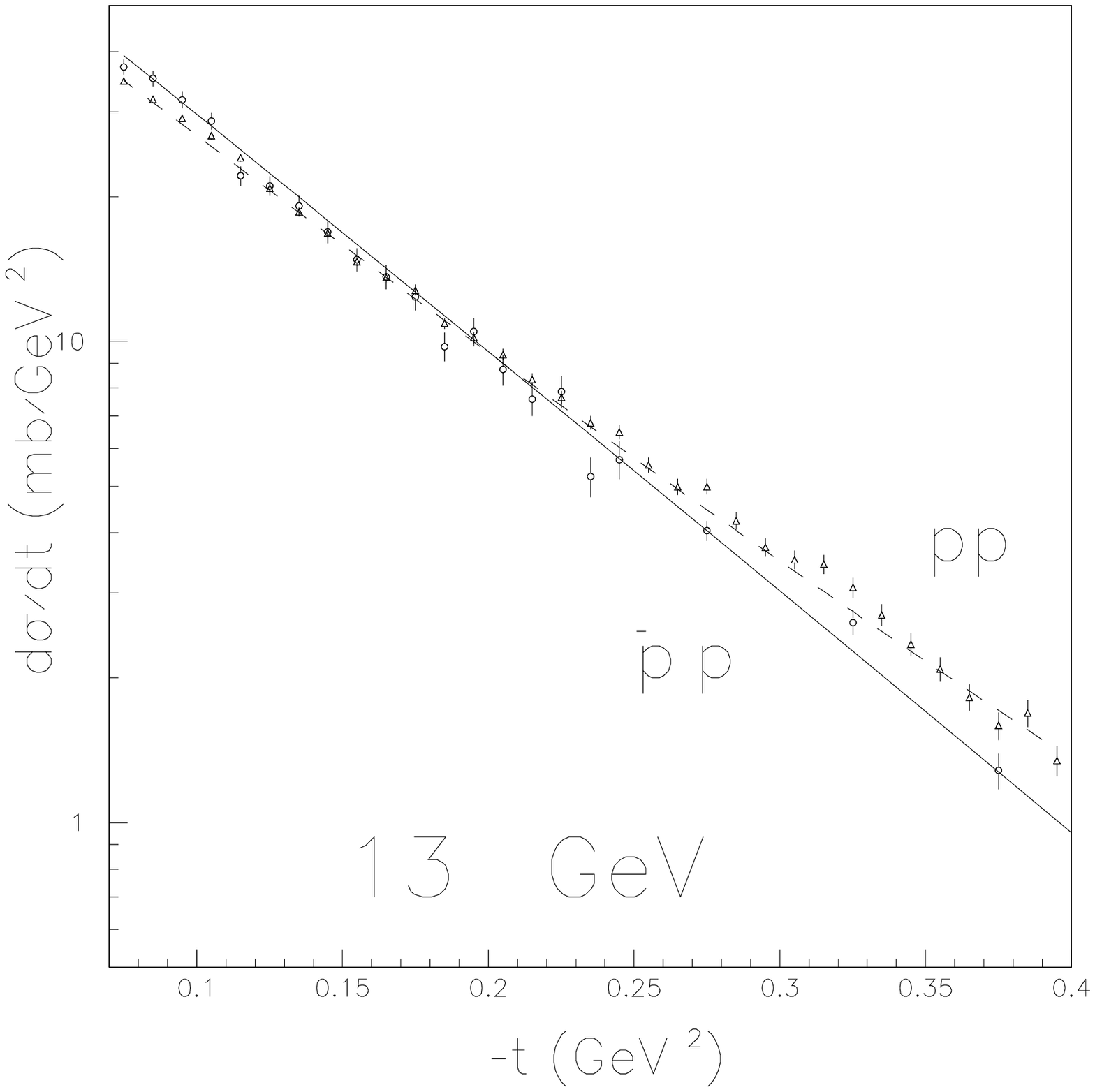}}
\epsfxsize=16 true cm
\centerline{\epsfbox{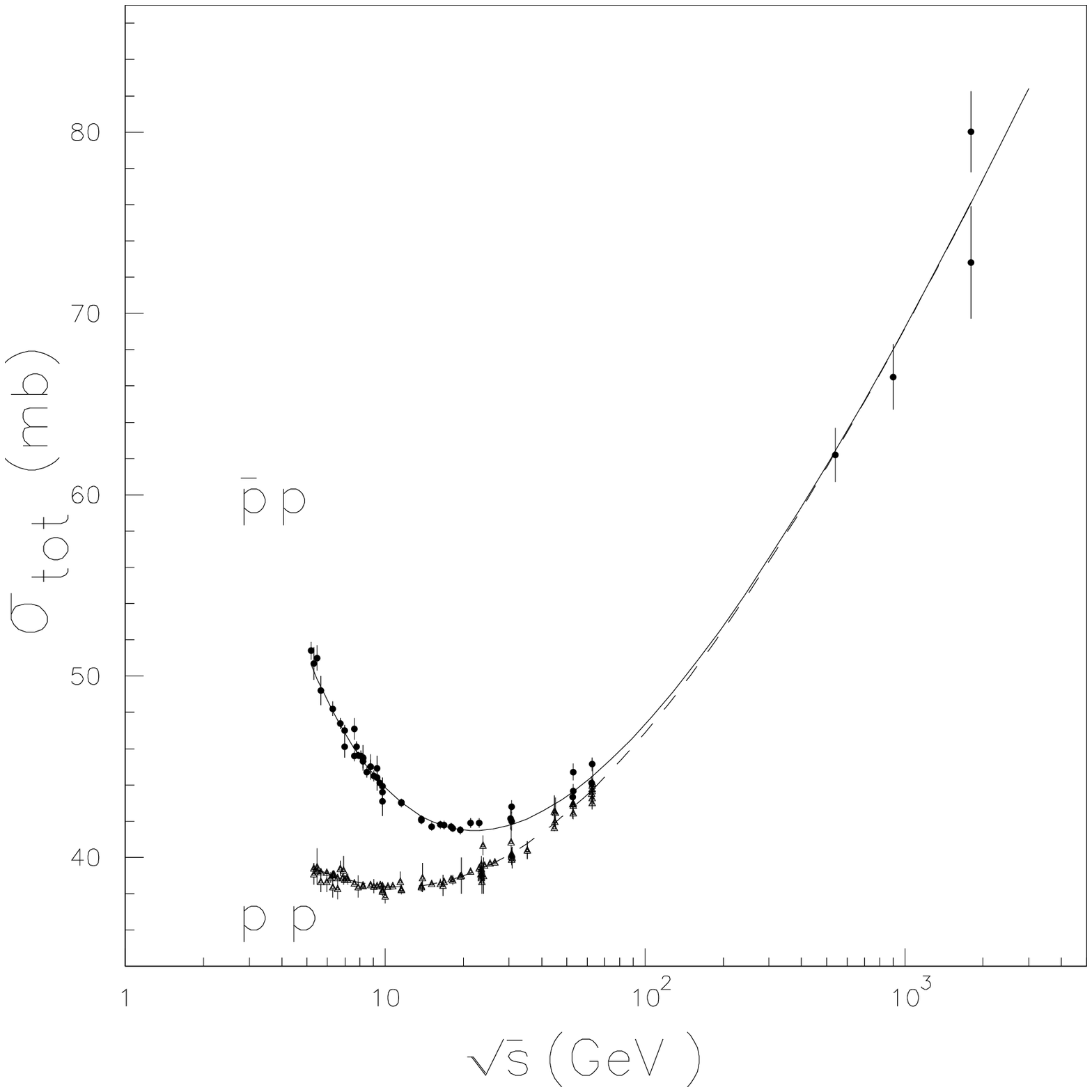}}
\epsfxsize=16 true cm
\centerline{\epsfbox{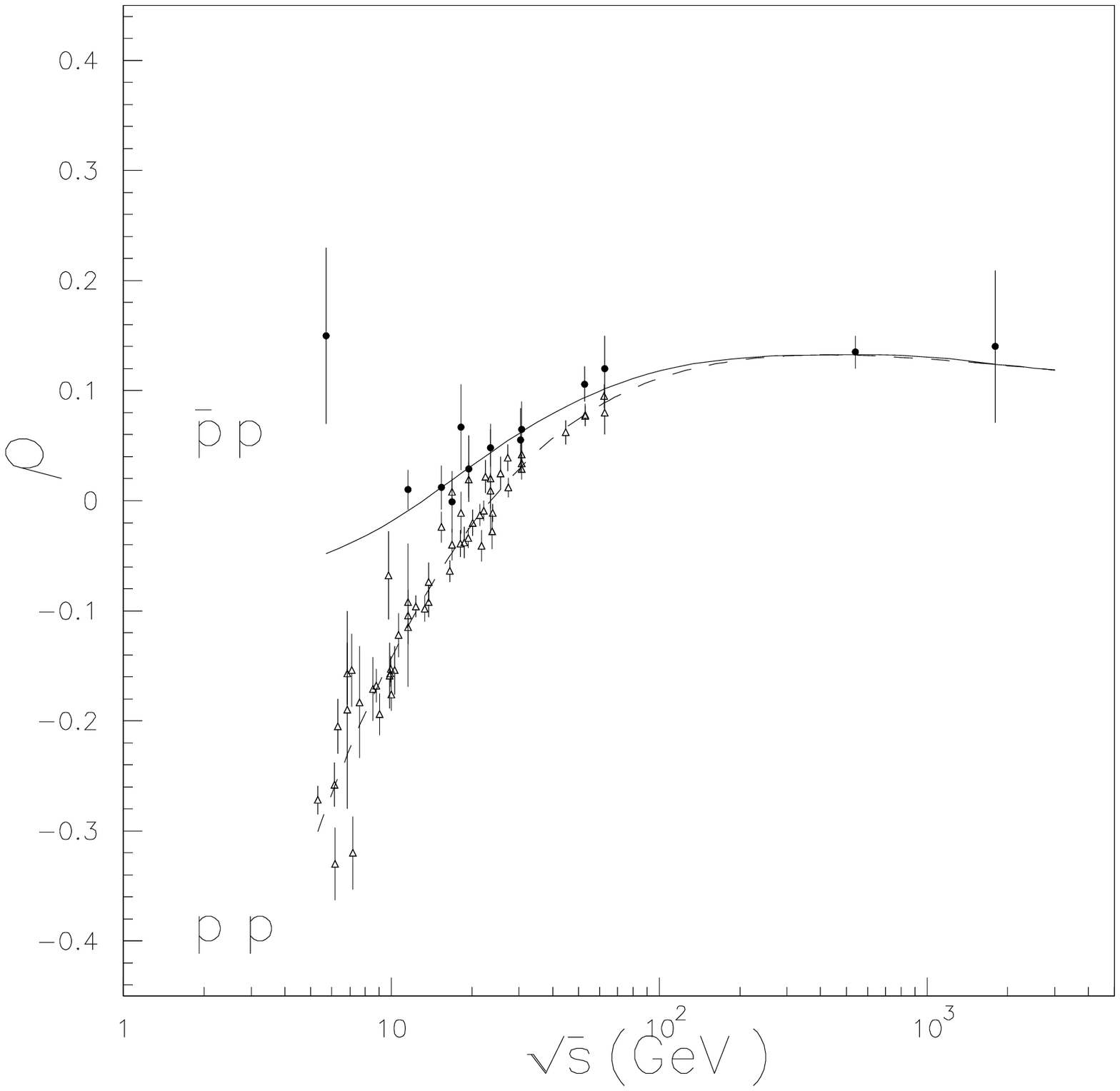}}
\epsfxsize=16 true cm
\centerline{\epsfbox{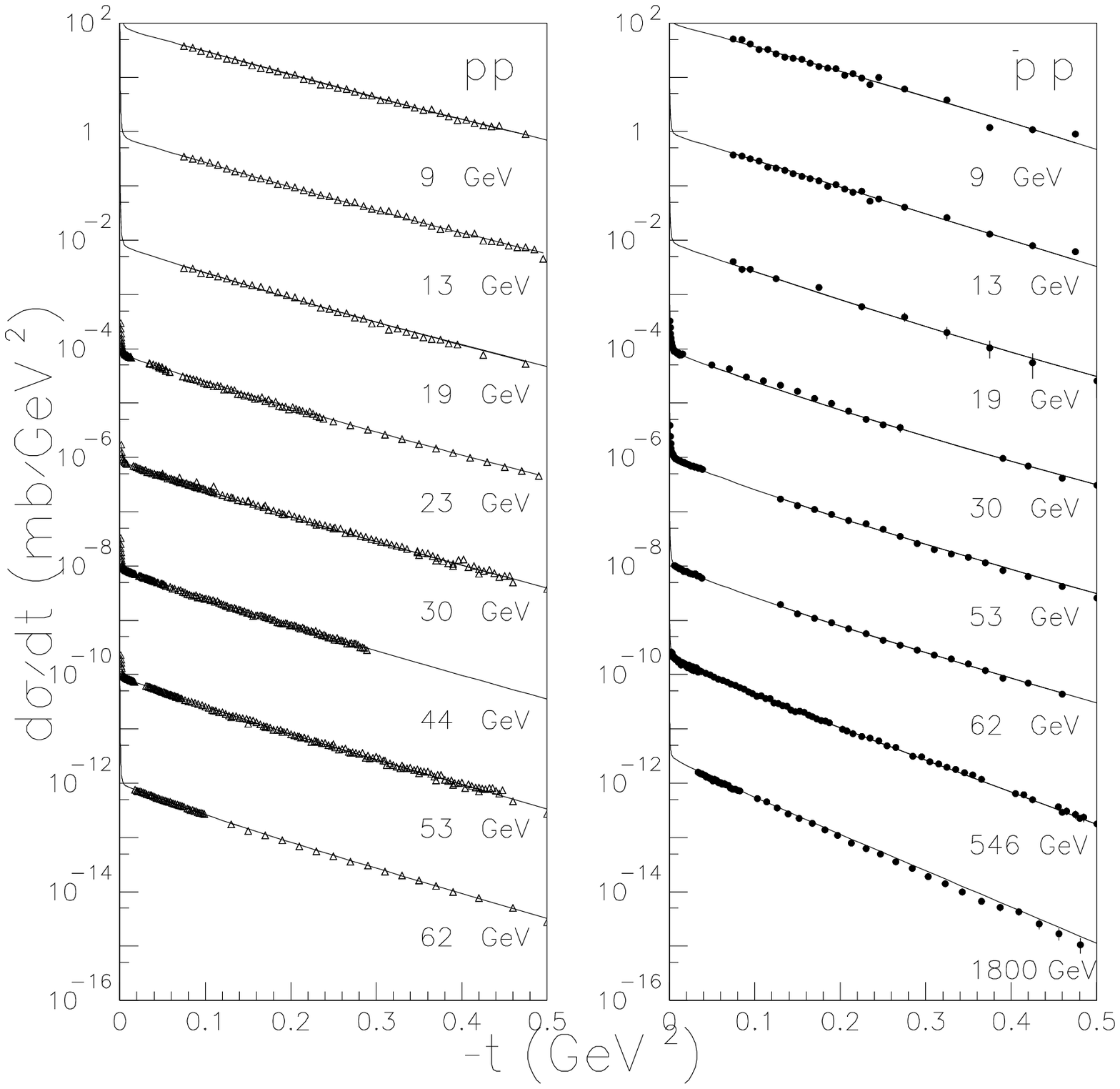}}
\epsfxsize=16 true cm
\centerline{\epsfbox{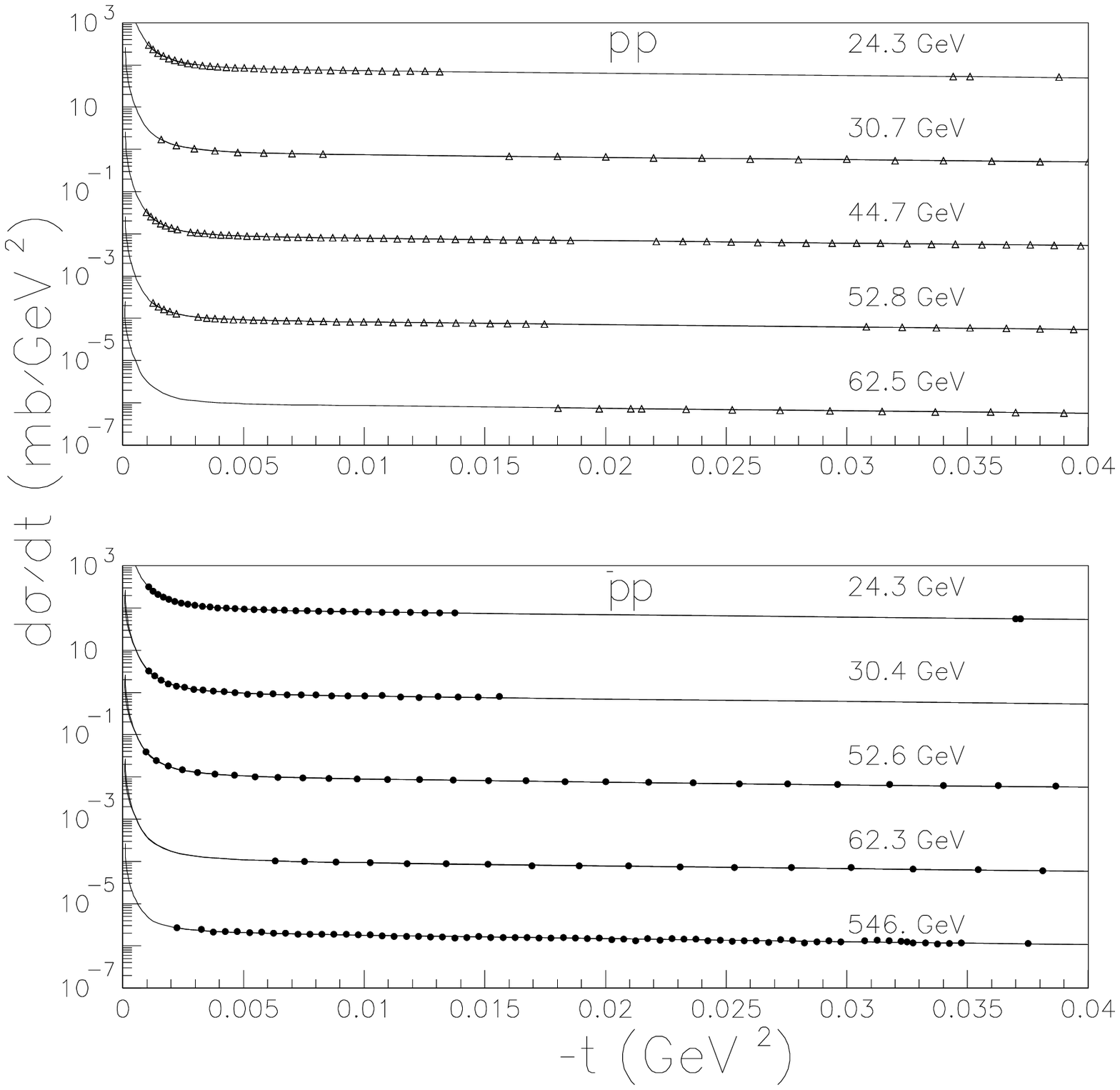}}
\epsfxsize=16 true cm
\centerline{\epsfbox{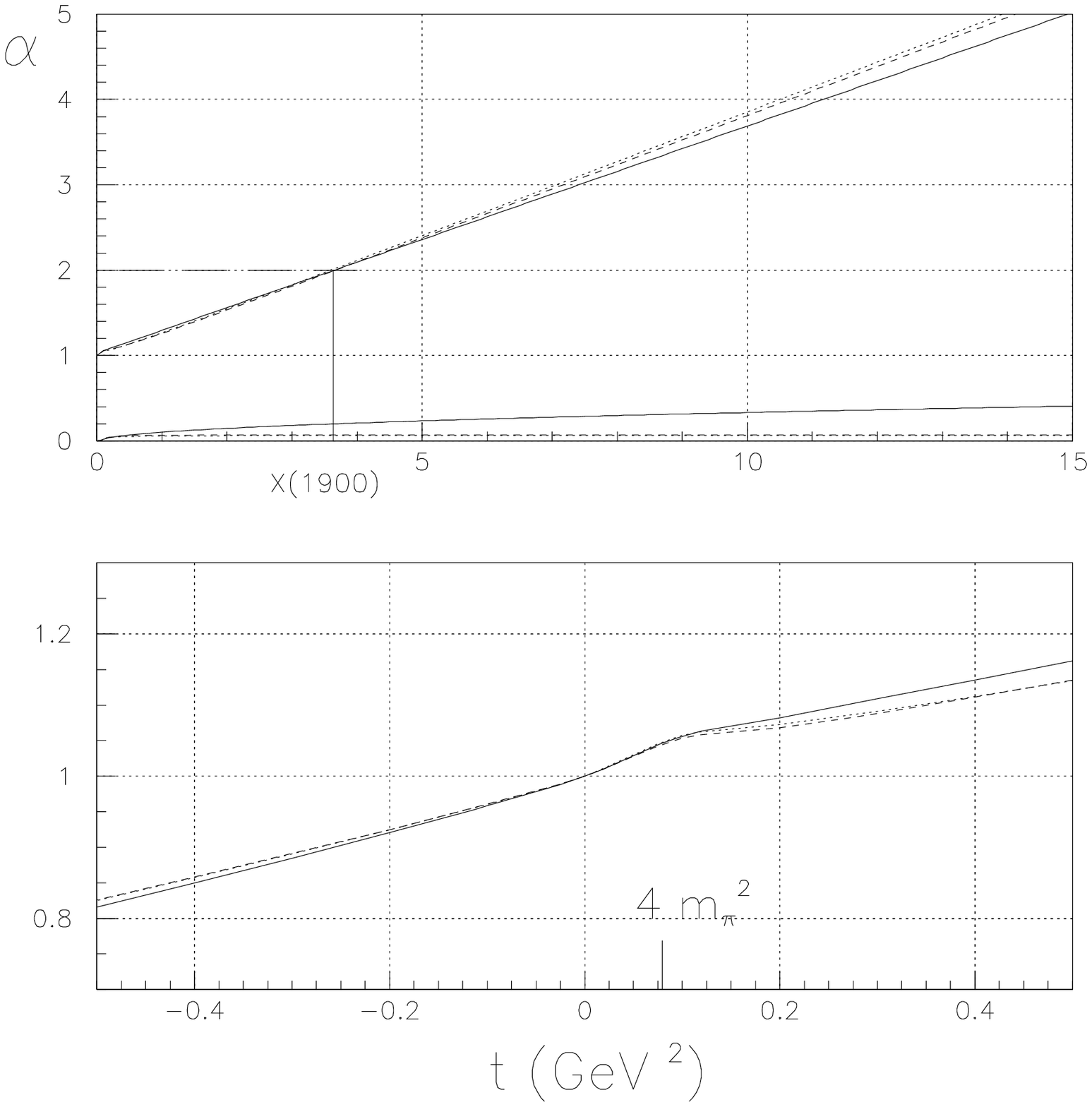}}
\bye